# Artificial Intelligence Probes for Interstellar Exploration and Colonization

Andreas M. Hein

Initiative for Interstellar Studies, 27-29 South Lambeth Road, London SW8 1SZ, United Kingdom

**Abstract**

A recurring topic in interstellar exploration and the search for extraterrestrial intelligence (SETI) is the role of artificial intelligence. More precisely, these are programs or devices that are capable of performing cognitive tasks that have been previously associated with humans such as image recognition, reasoning, decision-making etc. Such systems are likely to play an important role in future deep space missions, notably interstellar exploration, where the spacecraft needs to act autonomously. This article explores the drivers for an interstellar mission with a computation-heavy payload and provides an outline of a spacecraft and mission architecture that supports such a payload. Based on existing technologies and extrapolations of current trends, it is shown that AI spacecraft development and operation will be constrained and driven by three aspects: power requirements for the payload, power generation capabilities, and heat rejection capabilities. A likely mission architecture for such a probe is to get into an orbit close to the star in order to generate maximum power for computational activities, and then to prepare for further exploration activities. Given current levels of increase in computational power, such a payload with a similar computational power as the human brain would have a mass of hundreds to dozens of tons in a 2050 – 2060 timeframe.

## 1. Introduction

A recurring topic in interstellar exploration and the search for extraterrestrial intelligence (SETI) is the role of artificial intelligence (Armstrong and Sandberg, 2013; Dick, 2003; Tipler and Barrow, 1986; Tipler, 1994). On the one hand, interstellar exploration requires high levels of autonomy, if the probe is performing sophisticated scientific exploration of the target star system and by using more advanced technology, developing an infrastructure for future colonization (Hein, 2014a, 2014b). On the other hand, SETI-oriented publications imagine that a technologically advanced civilization will at some point likely transcend its biological existence and continue to exist and evolve in an artificial substrate. Colonizing the universe might then be much easier than within the confines of biological existence, as the body does not need to be sustained when the spacecraft is crossing the interstellar sea. In order to be considered as a form of colonization, we expect that some form of replication and intelligence is present in the colonization process. In the following, I call such an artificial life form with an intelligence that is superior to humans today "artificial general intelligence" (AGI) and any algorithm or computing device that exhibits cognitive features "artificial intelligence" (AI).

Past publications on interstellar exploration and SETI have mostly dealt with the principle feasibility of interstellar probes with a computationally-intense payload containing an AI/AGI. For example, Tipler describes the principle viability of mind-uploading into an artificial substrate, how a Daedalus-type interstellar probe could transport the AGI to other stars and gradually colonize the universe (Tipler and Barrow, 1986). Ray Kurzweil in "The Singularity is Near" describes nano-probes with AGI payloads that could even traverse small worm holes for colonizing the universe (Kurzweil, 2005). Finally Bradbury introduces the concept of "Matrioshka Brain" where a large number of spacecraft, powering AGIs orbit a star (Bradbury, 2001). Bradbury envisages that whole layers of orbital rings around a star could be created that harness the energy of a star. Hence, similar to a Dyson Sphere, SETI researchers could look out for the infrared signature of Matrioshka Brains for identifying extraterrestrial intelligences (Bostrom, 2003). Besides implications of AGI on SETI, Hein introduces several potential mission architectures based on AI or AGI interstellar probes (Hein, 2014a, 2014b). Various mission architectures are proposed, where AI/AGI probes pave the way for human interstellar colonization by creating space or surface colonies in advance to their arrival. Another proposed option is to use AGI to raise humans at another star, in order to enable transporting individual human cells to the stars and thereby avoid transporting the transportation of grown humans (Crowl et al., 2012).

However, apart from these high level description of AI/AGI probes, there is little research on the concept of an AI/AGI probe. By concept, I mean the spacecraft's functions, its subsystems, along with the mission architecture. In the following, I extrapolate from existing technology and trends to propose a concept for an AI/AGI probe along with an estimation of its mass. I demonstrate how the computing payload drives the design of the spacecraft. The conclusions from this exercise in explorative engineering (Drexler, 1991) are sufficiently general to be valid for different interstellar spacecraft that require large amounts of power for computing.

## 2. Literature survey

In a first step, I have to clarify, what I mean by AI and AGI. There is a common distinction between special AI and AGI. Special AI already exists today and is AI that can solve narrow tasks. Examples are chess, text mining, and expert systems. In these areas, AI has already achieved performances superior to humans. The landmark achievement of IBM's Deep Blue beating the then acting world chess champion is one of the most salient events in the history of AI. Recent achievements in image and location retrieval have demonstrated the superior performance of AI in these narrow tasks. By contrast, when I talk about AGI, I mean an AI that is capable of solving problems that are "general". Shanahan describes some of the characteristics of such general problems such as problems requiring some model of our world, requiring common sense and creativity (Shanahan, 2015). Common sense and creativity are considered the basic elements of an AGI. An AGI is required for problems that are "AI-complete" and can only be solved by an AI that exhibits human level intelligence (Mueller, 1987; Yampolskiy, 2012). However, the definition of AI-complete is itself subject to changes, as tasks such as human-level image recognition that have been considered AI-complete in the past can be performed better than humans today, without an AGI (Yampolskiy, 2012).

Today, we are still decades from an AGI (Bostrom, 1998) and there are proponents that consider AGI as infeasible such as Penrose (1999). One of the major tenets of classic AGI is that intelligence along with consciousness are properties that are inherently computable, i.e. can be captured by a program that is independent of its substrate. The "program" can be interpreted as the software and the substrate or embodiment as the hardware. This distinction harkens back to the "soul" and "body" distinction going back to Aristotle. This independence of the program from its embodiment is the so-called functionalist or cognitivist view of AGI. If it is true, approaches such as mind-uploading, i.e. changing the substrate on which the "mind" is running are feasible. However, this view has been repeatedly challenged by philosophers (Marchal, 1998, 1991, 1990; Preston and Bishop, 2002; Searle, 1990) and AI researchers (Pfeifer and Scheier, 2001).

There are conceptual and practical issues to overcome to create an AGI. The conceptual issues are basically questions about the nature of consciousness, which are still hotly debated. One of these assumptions is computability of human consciousness. It assumes that the human brain is based on computational principles. If yes, it should be replicable on an artificial substrate. If not, we may have to modify our models of computability. The practical issues are considered with the development of algorithms.

For the rest of the article, the distinction between AI and AGI is only in so far relevant, as I assume that the AI for such a mission needs computational power that is roughly equivalent to the computing power of the human brain. Hence, I assume that for both cases large computational power is required. The capabilities of the AI / AGI very much depend on the specific objectives of the mission. One can hypothesize that exploring the star system and collecting and analyzing data, or constructing an infrastructure from in-situ resources requires an AI beyond what exists today. These problems might be AI-complete and in such a case an AGI would be required.

## 3. Virtues of AGI-based interstellar travel

The main advantage of using an AI/AGI for interstellar travel is its hypothesized durability and lower mass, compared to transporting humans at similar or even superior capabilities than humans (Hein, 2014b). According to Hein (2014b), transporting humans requires a large habitat and environmental support systems, whereas an AGI-based mission would essentially require only a number of large computers. I say "hypothesized", as it is unclear if the mass for a computer that is able to store and run an AGI is actually much lighter than the equivalent of transporting a human to the stars. The threshold value is probably somewhere around 100 t per human (Hein et al., 2012; Matloff, 2006). According to Koomey's law (Koomey et al., 2011), the number of computations per joule of power dissipated has doubled every 1.57 years. Assuming that this trend will continue for the next 50 years, I will show at a later point that a breakeven point is going to be reached within this century.

However, even without the development of an AGI, pondering on an interstellar probe using some form of advanced AI has its merits. For example, we can imagine an AI that has been trained for exploring new environments and finding patterns that allows it to explore environmental features that potentially have a high scientific value. Such an AI might also require a high computational power. Hence, most of the conclusions I draw in the following remain valid independently of the precise form of AI.

## 4. Mission Objectives and Ground Rules

The existing literature on interstellar travel is not very specific about how spacecraft with a computation-heavy payload would look like. I define the following objectives for such a mission, where I assume that the first objective will be almost certainly part of such a mission and the latter two are more speculative:

- *Exploration of another star system:* The main objective of the mission is to explore the star system by collecting data, analyzing the data, and submitting analyzed data back to the Solar System;
- *Preparation for human colonization or continued exploration of other star systems:* These are ultimate objectives for a more extended AI-mission where I assume that the AI and spacecraft have considerable bootstrapping capabilities of using resources within the star system;
- *AGI-based colonization of star systems:* The AGI replicates and colonizes star systems, similar to a von Neumann probe (Freitas, 1980).

The nature and capabilities of the AI payload are not very important at this point. However, I assume that its cognitive and decision-making capabilities go far beyond what is currently possible.

Regarding the technology used on such a spacecraft, I make the following assumptions:

- *Computing technology:* I assume that the computing technologies are extrapolations of current technologies and extrapolating Moore's law several decades into the future is possible.
- *Power generation and heat rejection technologies:* I assume conservatively currently existing and near-future power generation and heat rejection technologies.

Hence, the AI probe design I present in this paper is rather a conservative AI probe and radically new technologies could enable AI probe that are much smaller for the same capability.

## 5. AI probe types

There is a long-lasting debate over the nature of intelligence, notably between proponents of Good Old-Fashioned Artificial Intelligence (GOFAI) (Haugeland, 1989) and approaches such as the "embodied cognitive science" approach (Pfeifer and Scheier, 2001). GOFAI is mainly concerned with classic approaches to AI that are based on symbolic manipulations. That is to say, intelligence is nothing else than the manipulation of symbols by a program: "A physical symbol system has the necessary and sufficient means for general intelligent action." (Newell and Simon, 1976, p.116) In other words, the "hardware" on which the program runs is irrelevant for intelligence. It is only the way the program is written that is important.

However, this reductionist perspective on AI has drawn considerable criticism in the past, most prominently from John Searle and his Chinese Room paradox (Searle, 1990, 1980). The main criticism of GOFAI is that it does not take aspects into consideration that play a key role in human intelligence such as (Pfeifer and Scheier, 2001):

- *Symbol grounding:* Without the relationship between symbols in a symbol manipulation system and reality, what the symbols actually stand for is always dependent on the interpretation of an external agent, i.e. a human.
- *Embodiment and situatedness:* Only when algorithms as "embodied" in the form of robots and are able to perceive their environment, they are able to interact with it.

Recent advances in the understanding of the human brain also contradict the GOFAI perspective on intelligence. The structure of the brain, i.e. the architecture of the hardware, seems to play an essential role in the way how the brain computes.

The practical consequences of this philosophical debate is very concrete. If AI can be reduced to a symbol system (program), intelligence can be copied, uploaded, etc. as it can be done for any computer program in principle such as hypothesized for whole brain emulation. However, if it is not a program, the "hardware" architecture and perception become important concerns.

Based on these AI paradigms, four types of AI can be defined that could be implemented in an AI probe:

- *Specialized AI probe:* Specialized AI. This type of AI would be based algorithms that exist today or in the near future such as GOFAI expert systems or neural networks. However, the AI is not capable of replicating the full breadth of human-level intelligence.

- *Whole brain emulation probe:* Replicas of the human brain are part of the AI payload of the spacecraft. "Whole brain emulation" can be defined as "the possible future one-to-one modelling of the function of the human brain" (Sandberg and Bostrom, 2008, p.1). The sensory inputs and motoric outputs of the emulated brain could be provided by a sophisticated virtual environment, the spacecraft's sensors, and virtual manipulators or robotic manipulators to achieve embeddedness and situatedness.
- *"From scratch" AI probe:* According to (Bostrom, 2014), this is a more speculative path to AGI, where the AGI does not resemble the human brain but is deliberately designed. How this feat could be accomplished is currently unknown.
- *Hive-mind AI probe:* Multiple independent agents collaborate or compete in order to exhibit a collective behavior. Agents could be any of the above forms of AI. Today, agents are frequently used in robotics to generate collective behavior.

AI probes can also be distinguished with respect to their objectives. It could be classic exploration where AI only serves as a means for realizing autonomous exploration of a star system, or could be more sophisticated such as preparing an infrastructure for human colonization or even entirely AI-based colonization where computational power is increased.

## 6. Mission Architectures

The mission architecture of an AI-based interstellar mission is based on a rather classic mission sequence for interstellar exploration as has been presented in Hein (2014b):

a. Launch
b. Cruise
c. Deceleration
d. Initial close-in to the star
e. Operate AI payload: The AI payload which is assumed to be power-intensive can only be operated to its full capacity within the target star system. Operating the AI payload allows for sophisticated data collection, analysis, and decision-making within the star system to devise further actions;
f. First observation and devising exploration strategy;
g. Start of exploration and replication: In order to start extensive exploration of the star system, the probe could either rely on already existing capabilities such as sub-probes or needs to manufacture the means for exploration within the star system. Some form of replication mechanism needs to be implemented in order to enable the construction of larger infrastructures such as communication antennas and propulsion systems. I assume that based on reasonable extrapolation of technologies currently under development such as 3D-printing of large truss structures in space that this is within reach.
h. Manufacturing space colonies or development of spacecraft for further exploration: For preparing long-term exploration goals such as colonization and launching spacecraft to other star systems, more sophisticated manufacturing capabilities are required, such as described in (Freitas and Gilbreath, 1982; Freitas, 1980; Jr and Zachary, 1981)

Fig. 1 shows a potential mission architecture for an AI probe mission. A possible propulsion system could be a laser sail (Forward, 1984; Hein et al., 2016; Lubin, 2016) along with a magnetic and electric sail deceleration system (Perakis and Hein, 2016).

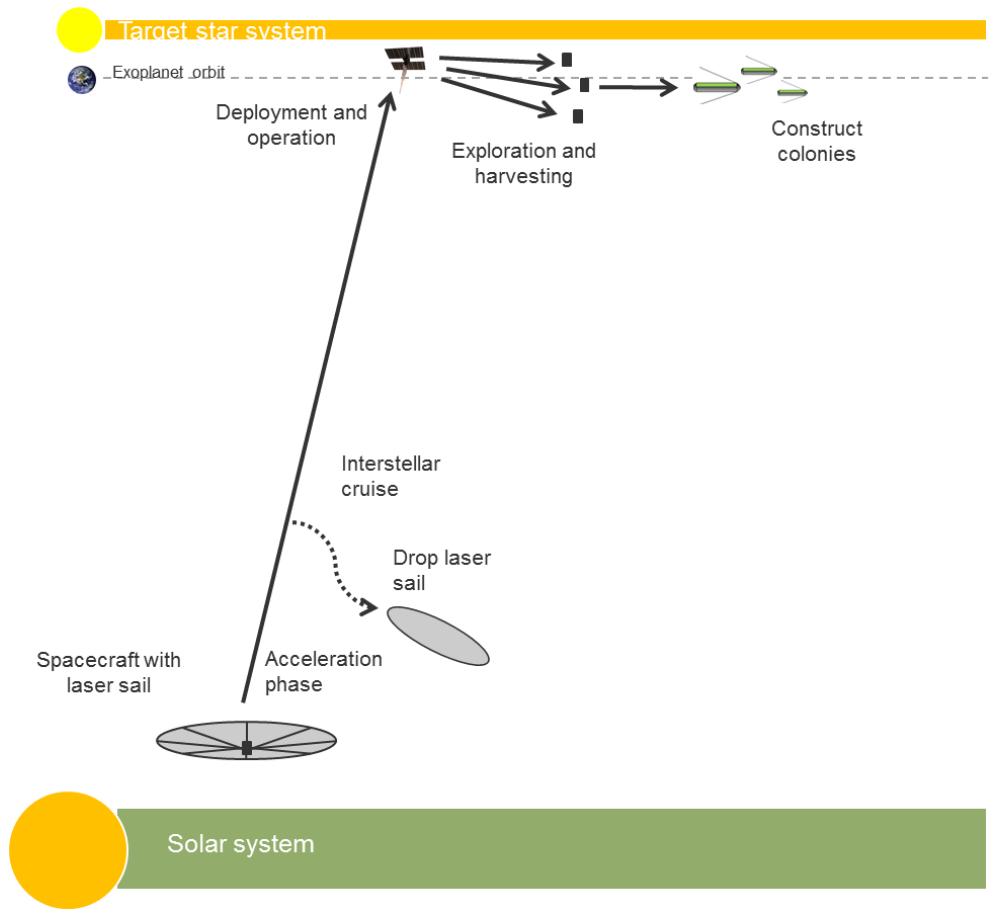

*Fig. 1: Potential mission architecture for an AI probe*

## 7. AI Probe Functions, Subsystems, and Design Drivers

The principal functions of the probe are listed below:

- Collect data from the star system via instruments;
- Store AI during travel via data storage
- Operate AI after arrival by using energy supply, data storage, and processing power
- Optional: Provide basis for self-sustainable growth of a civilization within the target star system, via:
    - Resource harvesting
    - Colony build-up
    - Interaction between AI and environment: robotics, manipulators, etc.
    - Build-up of an infrastructure by a self-replicating supply chain

Power is the basic requirement for information processing. The human brain needs about 25W for its operation (Kandel et al., 2000). Existing super computers have a much higher power consumption which is in the range of several hundreds of MW, as shown in Fig. 2.

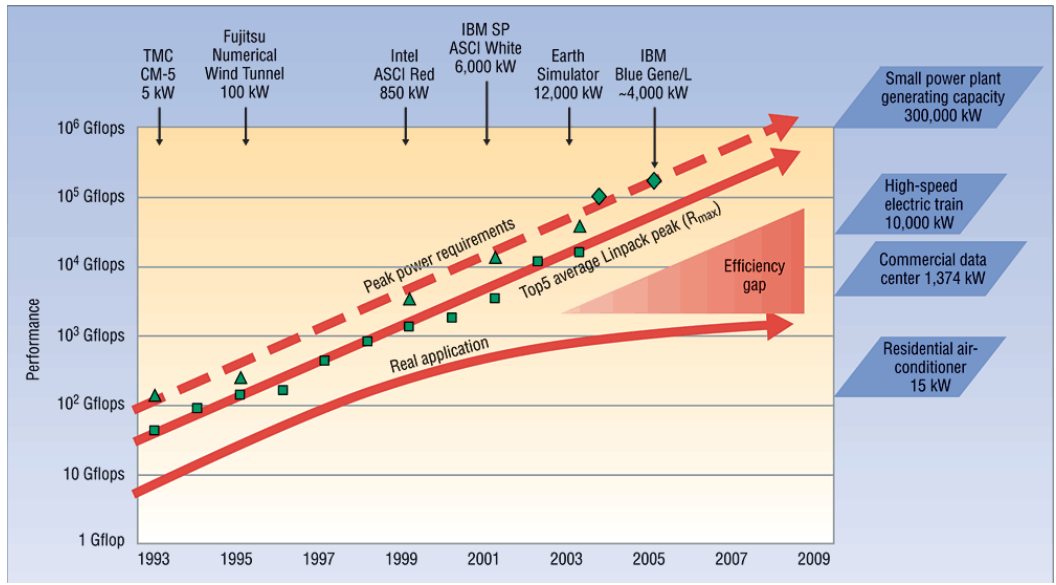

*Fig. 2: Performance and power requirements for supercomputers*
*https://www.computer.org/csdl/mags/co/2007/12/mco2007120050-abs.html*

Others have estimated the computing power of the human brain with $10^{20}$ flops (Sandberg and Bostrom, 2008). I will use this figure as a benchmark. The power required for this computation is estimated to be between 1MW ($10^6$ W) and 100GW ($10^{11}$ W), depending on how far current increases in computational power can be extrapolated. These figures somehow give us lower and upper bounds for the power requirements to simulate a human brain, which is on the order of 10W to $10^{11}$ W. The corresponding power generation systems would be equivalent to the power generation subsystems for a 3U-CubeSat and a hundred solar power satellite (Mankins, 2012). One can argue that existing computer architectures are very inefficient in replicating the function of a human brain. Future computer architectures or working principles of computers such as quantum computing could have a disruptive effect on power consumption (Markov, 2014). Hence, I keep the power consumption of a few dozen to hundreds of Watts as a lower boundary in case revolutionary new ways are found to reproduce the function of the human brain. However, a more conservative estimate would put the required power at dozens to hundreds of MW and lower values for AI that is less sophisticated.

If we assume that computation is running within a rather small volume in order to minimize the time the signal takes between computational nodes, we can expect that large amounts of heat are generated in a small volume. Heat rejection is currently one of the main challenges of super computers (Nakayama, 2014). The currently prevalent approach for heat rejection in super computers are heat pipes that transport a cooling liquid to the processors and the heated liquid away from them. The current heat density is as high as 10 kW/cm² for super computers, which is about one order of magnitude higher than the heat density inside a rocket engine nozzle (Nakayama, 2014) . Contrary to terrestrial heat rejection approaches, in space, heat can only be rejected without mass loss via radiation, which requires large surface areas that are faced towards free space. Advanced radiators could reject about 1kWt/kg thermal power per kilogram in the near future (Adams et al., 2003; Hyers et al., 2012; Juhasz and Peterson, 1994). Consequently, about 100 tons of radiator mass would be required for rejecting 100MW and 1 ton for 1MW respectively.

Regarding the mass of the computing unit, current on-board data handling systems (OBDH) have a computing power on the order of 100 DMIPS per kg, as shown in Fig. 3 as orange points. Fig. 3 shows the evolution of DMIPS for processors in general. The spacecraft OBDH from the literature have DMIPS values that are about two orders of magnitude below the values for terrestrial processors. If we extrapolate these values to the 2050 timeframe, we can expect spacecraft OBDH with a processing power of 15 million DMIPS per kg. As DMIPS and flops are different performance measures, I use a value for flops per kg from an existing supercomputer (MareNostrum) and extrapolate this value (0.025Tflops/kg) into the future (2050). By 2050, I assume an improvement of computational power by a factor $10^5$, which yields $0.025*10^{17}$flops/kg. In order to achieve $10^{20}$flops, a mass of dozens to a hundred tons is needed. I assume an additional 100 tons for radiator mass and with 1kW/kg for the solar cells, about 100 tons for the solar cells. This yields a total mass for an AI probe on the order of hundreds of tons, which is roughly equivalent to the mass of the Daedalus spacecraft of 450 tons (Bond, 1978).

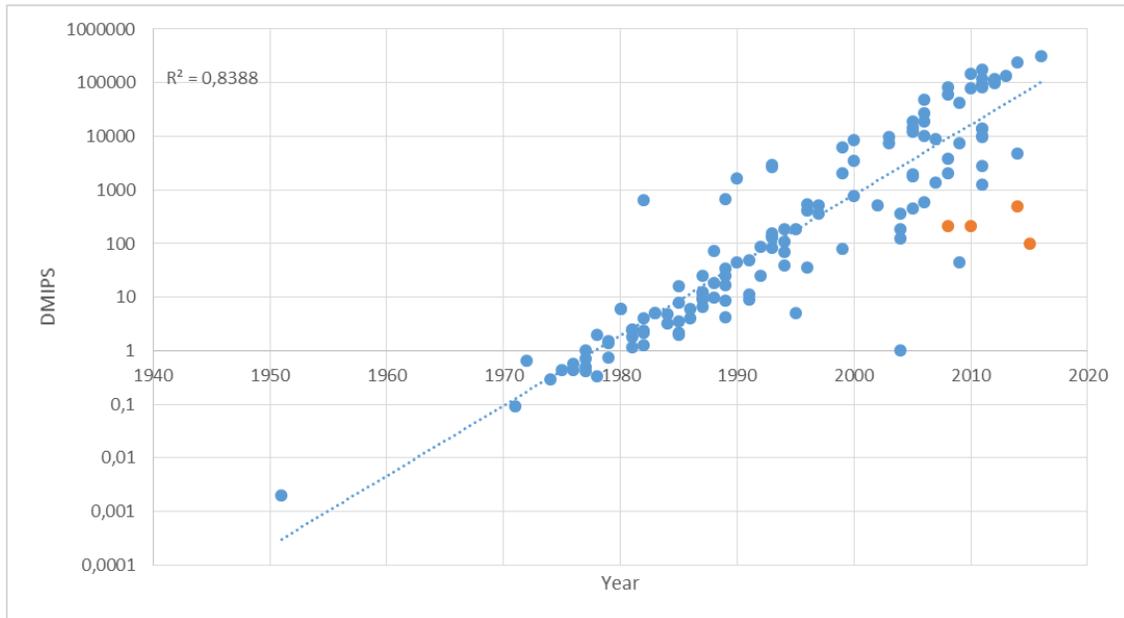

*Fig. 3: DMIPS versus year for processors in general and space on-board data handling systems (https://en.wikipedia.org/wiki/Instructions_per_second)*

Table 1 shows the mass estimates for the main spacecraft subsystems and its total mass. The mass estimate is only valid for the part of the spacecraft that actually arrives at the target star system.

*Table 1: Mass estimate for AI probe*

| Spacecraft subsystem | Specific mass | Subsystem mass [t] |
|---|---|---|
| *Computing payload* | $0.025*10^{17}$flops/kg | 40 |
| *Solar cells (current technology)* | 1kW/kg | 100 |
| *Radiators* | 1kWt/kg | 100 |
| *Other subsystems (50% of computing payload)* |  | 20 |
| **Total mass** |  | **260,000** |

With the assumptions introduced before, several conclusions can be drawn with regards to the architecture / configuration that are specific to an AI probe:

- *Large solar panels:* Unless other power sources such as nuclear power is used, the probe will depend on large solar panels / solar concentrators for generating power for the AI payload.
- *The spacecraft is at least initially operated close to the star.* In order to maximize power input from the star and to minimize solar power generator mass, the probe should be located as close to the star as possible. The minimum distance is constrained by the maximum acceptable temperature for the spacecraft subsystems and an eventual heat shield that protects against the starlight. For that purpose, it needs a star-shield to protect against the heat and radiation from the star. A trade-off between the heat shield mass and the mass savings from lower solar power generator mass needs to be made. However, note that the AI payload itself is a source of intense heat and more sensitive spacecraft subsystems such as sensors need to be located distant to the AI payload, e.g. on a boom.
- *AI payload switched off outside star system:* AI is switched off outside the target star system, as there is no power source available for its operation;
- *Large radiator:* A large radiator is needed for rejecting the heat generated by the AI payload;
- *Compact computing unit:* The computer is either super-compact or distributed. However, with a distributed system, communication speed becomes an issue and it is therefore likely that the architecture will be as compact as possible to minimize the time for signals to travel within the payload.

Fig. 4 and Fig. 5 show an artist's impression of an AI probe with its main subsystems. The design is similar to spacecraft with nuclear reactors. However, spacecraft with nuclear reactors do not have solar cells, as they are designed precisely for environments without sunlight.

The AI payload likely has a cylindrical shape, as it is easier for the heat rejection system to have one backbone heat channel and then smaller, radial pipes that reject heat from the processing units. The heat is rejected via large radiators. Fig. 7 shows that the radiator size decreases with distance from the payload, as less and less fluid is available for rejection. It is furthermore better to reject the heat quickly. Hence, the larger size of the radiators close to the payload. The radiator is perpendicular to the payload in order to avoid heat radiation from the payload being absorbed by the radiators. Fig. 5 shows an additional heat shield between the payload section and the radiators, in order to prevent radiative heat transfer from the payload.

In order to maximize energy intake from the star, the spacecraft is located as close as possible to the star. However, then, strong thermal radiation and particles from the star impact the spacecraft. In order to avoid heat and particle influx from the star, a heat and radiation shield for protecting the spacecraft from these particles. The shield is located in the direction of the star and shields the payload.

The spacecraft needs to be constantly maintained and parts replaced or repaired. This is similar to existing terrestrial supercomputers. There is no way that a system of this complexity does not need repair. If the computer is modular, these modules are replaced on a regular basis and parts replaced within these modules. I can imagine a storage depot of parts and robots that replace these parts. With more advanced technology available, robots that reproduce even very complex replacement parts can be imagined.

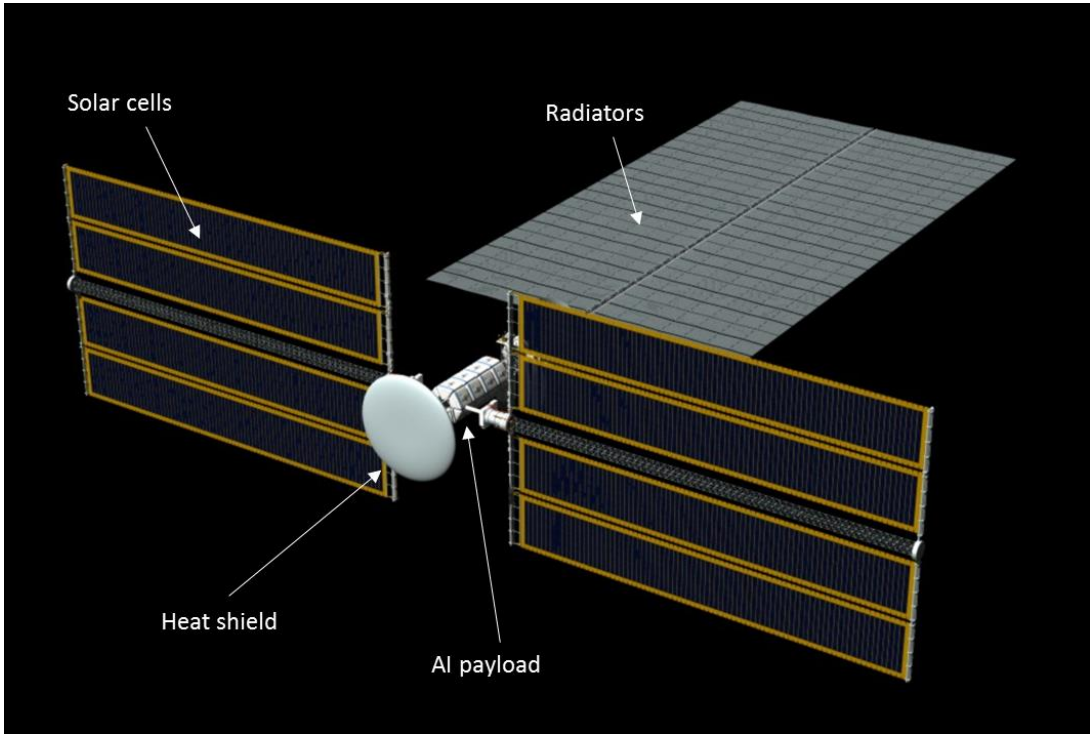

*Fig. 4:AI probe subsystems (Image: Adrian Mann)*

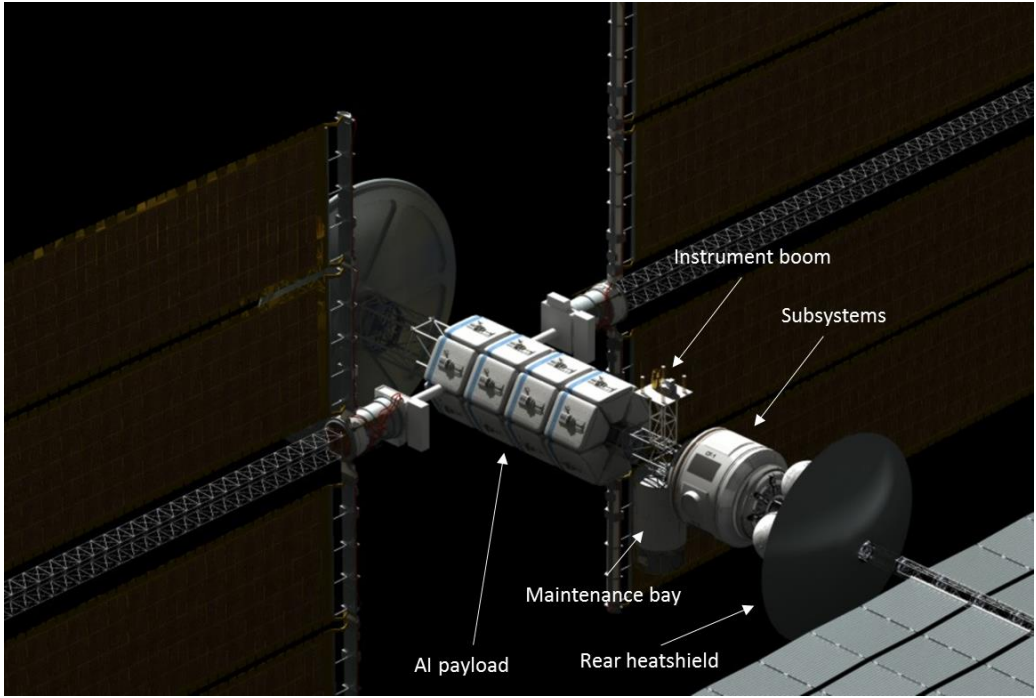

*Fig. 5: View from the back of AI probe (Image: Adrian Mann)*

Under the assumption that during the 2050 to 2070 timeframe, computing power per mass is still increasing by a factor of 20.5, it can be seen in Fig. 6 that the payload mass decreases to levels that can be transported by an interstellar spacecraft of the size of the Daedalus probe or smaller from 2050 onwards. Such a mission might be subject to the "waiting paradox", as the development of the payload might be postponed successively, as long as computing power increases and consequently launch cost decrease due to the lower payload mass. Furthermore, under the assumption that an advanced AI payload has equal capabilities for exploration than a human and the mass required for transporting a human over interstellar distances is about 100t (Hein et al., 2012; Matloff, 2006), the breakeven point for an AI probe compared to a human mission is somewhere between 2050 and 2060. Furthermore, it is obvious that an interstellar mission with a human crew cannot be downscaled indefinitely. Smith (2014) estimates that at least a population of 20,000 needs to be transported.

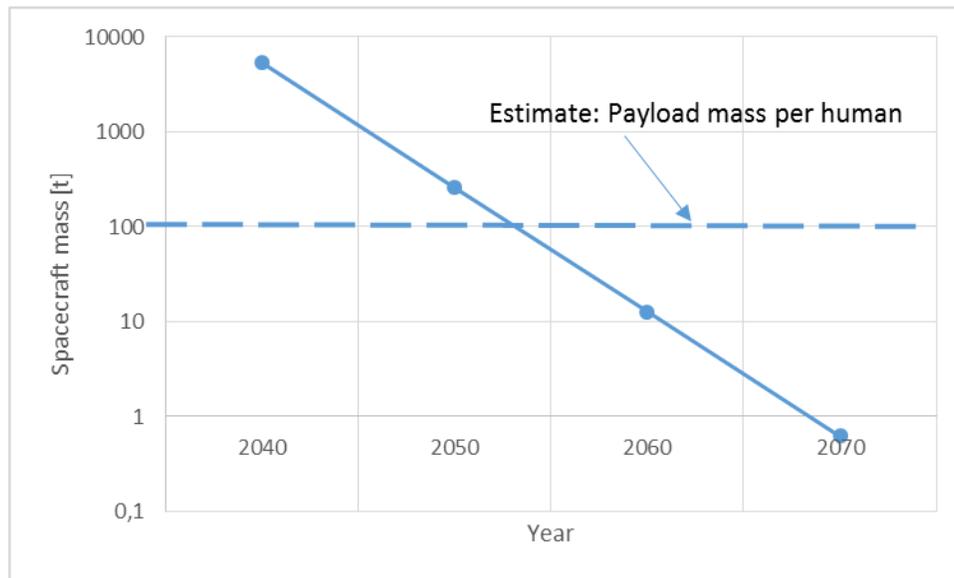

Fig. 6: Spacecraft payload mass vs. year of development

## 8. SETI implications

Up to this point I have presented the results for an AI probe that is based on existing technologies and extrapolation of current trends. One interesting question is, how computationally-intense probes of other species could look like. In such a case, these assumptions are not valid. Hence, I refer to physical limits to computational power from the literature and how that would impact the design of an AI probe.

Several physical limits to computation have been proposed in the literature (Bennett and Landauer, 1985; Lloyd, 2000; Markov, 2014) but the status of these limits is disputed (Markov, 2014). Nevertheless, these limits can be used to explore concepts for AI probes of a civilization that is more advanced than ours.

| Reference | Bound |
| --- | --- |
| Minimum operations per energy (Margolus and Levitin, 1998) | $6*10^{33}$ operations / s / joule |
| Minimum operations per mass (Bremermann, 1967, 1962) | $1.36*10^{50}$ bits / s / kg |
| Lower energy limit for one bit of memory manipulation (Landauer, 1961) | $2.87*10^{-23}$ J/bit |

It can easily be seen that these limits are numerous orders of magnitude larger than the values used previously. Using the energy limit from Margolus and Levitin (1998), a spacecraft using 1 W of power could perform computing operations of billions of human brain equivalents. However, an advanced AI probe may still maximize energy harvesting from a star and maximize mining resources for construction such as self-replication or a larger computational infrastructure. Hence, at least the power harvesting would take place as close to the star as possible, although I can imagine that the computing is performed at a place where heat can be radiated away to the cosmic background better than close to a star. For example, the power is beamed via a laser from the solar power collectors to the computing units. For gathering resources, celestial bodies in the star system are the natural choice. Based on this reasoning, the following conclusions can be drawn:

- Unless power sources such as fusion are used, power generation for computing is likely to take place close to the Sun to limit solar panel size;
- The computing units are likely to be shielded from the heat and radiation of the star and if possible placed at a more distant location from the star to maximize heat rejection efficiency and minimize heat shield size;
- If possible, probes will harvest material close to the star in order to minimize the time and energy for transportation.

From these design considerations, the following SETI observational conclusions can be drawn:

- Search for objects that reject heat in the infrared and are distant to the star could indicate computing units;
- Star occultation could hint at large-scale solar power generation, similar to hypothesized by Dyson spheres and Matrioshka Brains (Badescu, 1995; Bradbury, 2001);
- Mining strips could be on celestial bodies close to star.

Except for the search of objects emitting in the infrared far from the star, the two other conclusions are similar to the ones for Dyson spheres, Matrioshka Brains, and to the field of Search for Extraterrestrial Artifacts (SETA) in general (Freitas and Valdes, 1985; Freitas, 1983).

## 9. Conclusions

A concept for an interstellar probe with a computational-heavy payload was presented. Such a probe could perform the autonomous exploration of a star system and development of an infrastructure enabling for human colonization. AI spacecraft development and operation will be constrained and driven by three aspects: power requirements for the payload, power generation capabilities, and heat rejection capabilities. Based on the extrapolation of existing technologies and trends, I estimated that the payload of such an interstellar probe that has a similar computing power as the human brain is likely to have a mass of hundreds of tons in the 2050 timeframe and a mass of dozens of tons in the 2060 timeframe. Furthermore, I reasoned that a more advanced civilization that has used such probes based on much more advanced computing technology would likely to deploy the power generation infrastructure close to a star and the computing units protected from the star's radiation and facing free space to maximize heat rejection. Furthermore, resource harvesting is also likely to take place close to these infrastructures to minimize resource expenses for transportation. For future work, I propose a more detailed analysis of the spacecraft's architecture, its instruments, and specifics about the computing payload for assessing the scientific value of such a mission.

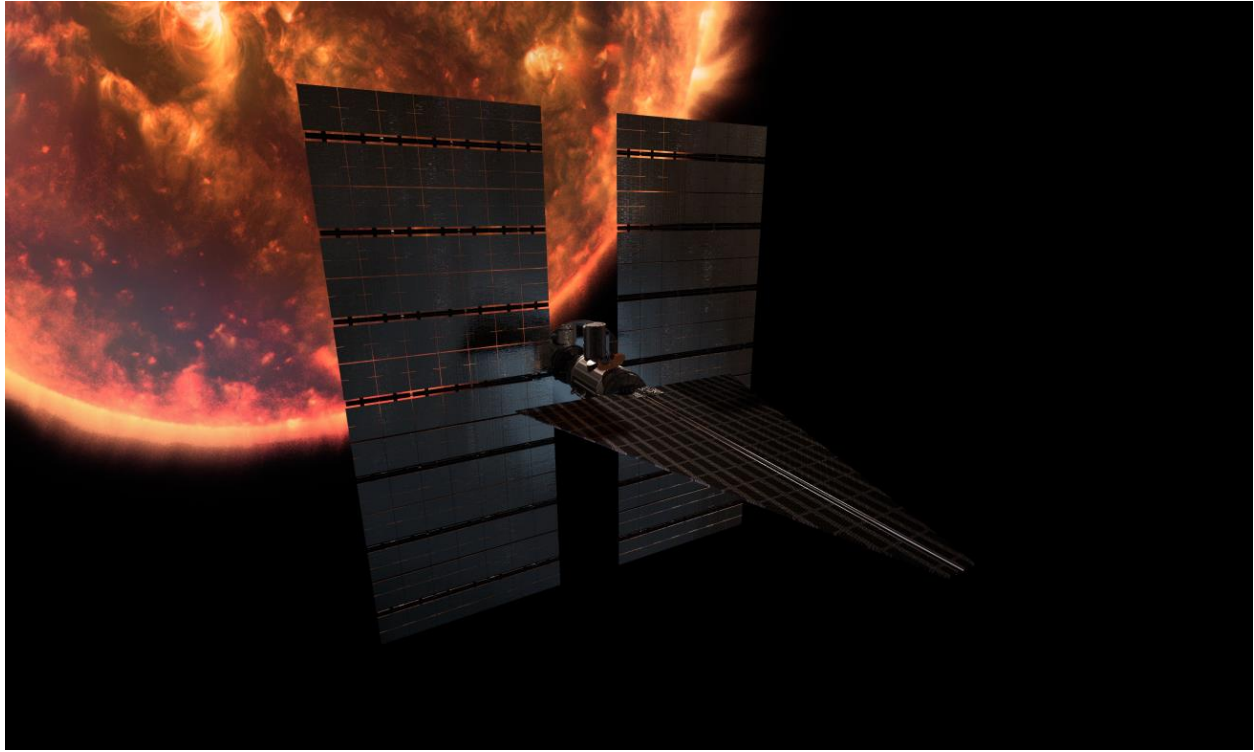

*Fig. 7: Artist impression of an AI probe close to a star (Image: Efflam Mercier)*